\documentclass[twocolumn,prc,showpacs,preprintnumbers,superscriptaddress,floatfix]{revtex4}
\usepackage{dcolumn}
\usepackage{bm}
\usepackage{longtable}
\usepackage{graphicx,epsfig,latexsym,amssymb}
\usepackage{multirow,amsmath,array,booktabs,color}
\usepackage[section]{placeins}

\begin{document}

\title{Research on the halo in $^{31}$Ne with complex momentum
representation method}
\author{Ya-Juan Tian}
\affiliation{School of physics and materials science, Anhui University, Hefei 230601,
P.R. China}
\author{Quan Liu}
\email[E-mail:]{quanliu@ahu.edu.cn}
\affiliation{School of physics and materials science, Anhui University, Hefei 230601,
P.R. China}
\author{Tai-Hua Heng}
\email[E-mail:]{hength@ahu.edu.cn}
\affiliation{School of physics and materials science, Anhui University, Hefei 230601,
P.R. China}
\author{Jian-You Guo}
\email[E-mail:]{jianyou@ahu.edu.cn}
\affiliation{School of physics and materials science, Anhui University, Hefei 230601,
P.R. China}
\date{\today }

\begin{abstract}
Halo is one of the most interesting phenomena in exotic nuclei especially
for $^{31}$Ne, which is deemed to be a halo nucleus formed by a $p-$wave
resonance. However, the theoretical calculations don't suggest a $p-$wave
resonance using the scattering phase shift approach or complex scaling
method. Here, we apply the complex momentum representation method to explore
resonances in $^{31}$Ne. We have calculated the single-particle energies for
bound and resonant states together with their evolutions with deformation.
The results show that the $p-$wave resonances appear clearly in the complex
momentum plane accompanied with the $p-f$ inversion in the single-particle
levels. As it happens the $p-f$ inversion, the calculated energy, width, and
occupation probabilities of major components in the level occupied by
valance neutron support a $p-$wave halo for $^{31}$Ne.
\end{abstract}

\maketitle

\section{Introduction}

Research on the nuclei far from the $\beta-$stability line, especially close
to the neutron drip line, is a hot topic in nuclear physics. These nuclei
with large ratio $N/Z$ compared to their isotopes may have new features: the
emergence of halo~\cite{Tanihata1985}, the disappearance of traditional
magic number and the generation of new magic number~\cite{Ozawa2000}, and so
on. One of the most interesting characteristics is the halo: one (or two)
weakly-bound valence nucleon(s) spatially extend(s) far beyond the nuclear
core by means of the quantum tunneling effect. This structure is ascribed to
an occupation of the $l=0$ or $l=1$ orbit by valence nucleon~\cite%
{Sagawa1992}. The first confirmed halo structure appears in $^{11}$Li, which
is considered as two valence neutrons occupying the $2s_{1/2}$ orbit~\cite%
{Meng1996}. Afterwards, more halo structures such as $s-$wave halo and $p-$%
wave halo, two-body halo and three-body halo, etc. were explored and
researched in experiments~\cite{Tanihata2013} and theories~\cite{Meng2015}.

In these researches on halo nuclei, $^{31}$Ne has received more attention
since it is located in a mixing region of normal and intruder shell
configurations. According to the shell model's scenarios, $^{31}$Ne ground
state should be a $^{30}$Ne core in its $0^{+}$ ground state with a $%
1f_{7/2} $ valence neutron. However, the calculations in Refs.~\cite%
{Poves1994,Descouvemont1999} favored this valence neutron populating in the $%
2p_{3/2}$ intruder orbit. The orbit with low one-neutron separation energy ($%
S_{n}\simeq 0.33$ MeV~\cite{Audi2003}) supports a halo structure. The
one-neutron removal cross sections measured at RIKEN~\cite{Nakamura2009}
suggested that $^{31}$Ne is a halo nucleus. The data can be most easily
interpreted as the ground state spin parity being $3/2^{-}$ rather than $%
7/2^{-}$. Further experiment~\cite{Takechi2012} showed the interaction cross
sections of $^{31}$Ne are significantly greater than those of their
neighboring nuclides, which cannot be explained by the single-particle shell
model under the assumption that the valence neutron occupies the $1f_{7/2}$
orbit.

In order to clarify the ground state spin-parity, some theoretical models
were introduced for $^{31}$Ne. In Ref.~\cite{Horiuchi2010}, the Glauber
model was used to analyze the one-neutron removal cross sections. The result
suggests a strong $2p_{3/2}$ configuration in the $^{31}$Ne ground state. In
the framework of the particle-rotor model, Urata $et$ $al.$ applied the
Glauber theory to calculate the reaction cross sections, and found that the
difference of the reaction cross sections between $^{30}$Ne and $^{31}$Ne is
much larger for the configuration with $\beta _{2}=0.2$ than that with $%
\beta _{2}=0.55$, leading to a consistent description of one-neutron removal
cross section with $\beta _{2}=0.2$~\cite{Urata2011,Urata2012}. Based on a
quantitative analysis of the reaction cross sections of $^{28-32}$Ne by $%
^{12}$C at 240 MeV/nucleon using the double-folding model, Minomo $et$ $al$.
concluded that neutron-rich Ne isotopes are strongly deformed and $^{31}$Ne
has a halo structure with the spin parity $3/2^{-}$~\cite%
{Minomo2011,Minomo2012}. Using the density obtained by antisymmetrized
molecular dynamics, the double folding model well reproduces the measured
interaction cross sections of $^{31}$Ne with the quadruple deformation $%
\beta_2\sim 0.4$~\cite{Sumi2012}.

In addition to the reaction cross section, the knowledge on the
single-particle resonances is critical to understand the halo structure of
exotic nuclei. In Ref.~\cite{Hamamoto2010}, the scattering phase shift
approach was applied to explore the single-particle resonances in $^{31}$Ne
in a deformed mean field. The observed large Coulomb breakup cross section
of $^{31}$Ne can be interpreted in terms of $p-$wave neutron halo together
with the deformed core $^{30}$Ne. With the analytic continuation approach
for resonances, $^{31}$Ne is predicted to possess a $p-$orbit one-neutron
halo structure based on the criteria of a large matter radius, low
separation energy, and large population of a low$-l$ orbit~\cite{Zhang2014}.
The halo structure of $^{31}$Ne was also claimed from an occupation of the
weakly bound level 1/2[330] by valance neutron in the complex scaling
calculations~\cite{Liu2012}.

All these are inclined to think that $^{31}$Ne is a deformed halo nucleus
with the spin parity $3/2^{-}$. However, we don't know that the $3/2^{-}$
comes from the $2p_{3/2}$ or $1f_{7/2}$ orbit and whether it happens the $%
p-f $ inversion in $^{31}$Ne. Since the $2p_{3/2}$ and $1f_{7/2}$ orbits in $%
^{31}$Ne have positive energy, it is necessary to treat the resonant states.
The calculations in Ref.~\cite{Hamamoto2010} exposed the $1f_{7/2}$
resonance, but the $2p_{3/2}$ resonance is missed. Although more resonant
states are obtained in the complex scaling calculations~\cite{Liu2012}, the $%
2p_{3/2}$ resonance has not been discovered. The reason may be that the $%
2p_{3/2}$ state corresponds to a broad resonance, while broad resonance is
difficult to obtain in theory. Recently, we have developed a complex
momentum representation (CMR) method for resonances~\cite{li2016,Fang2017},
which is effective for not only narrow resonances but also broad resonances
with many advantages in comparison with some other methods~\cite%
{li2016,Tian2017}.

In the present paper we apply the complex momentum representation method to
explore the resonant states in $^{31}$Ne. The missed $2p_{3/2}$ resonance is
obtained in accompanying with the $1f_{7/2}$ resonance. The $p-f$ inversion
is exposed clearly in the single particle levels. Moreover, we have
calculated the density distributions and the occupation probabilities of
major components, which support a $p-$wave halo nucleus for $^{31}$Ne. In
the following, we express the theoretical formalism in Sec. II. In Sec. III,
we present numerical details and results. A summary is collected in Sec. IV.

\section{Formalism}

To explore the resonances in $^{31}$Ne, we first present the theoretical
formalism of complex momentum representation method with the Hamiltonian as
\begin{equation}
H=T+V,  \label{Hamiltonian}
\end{equation}%
where $T$ is the kinetic operator and $V$ is the interaction potential. The
kinetic energy operator $T=p^{2}/2m$ with the reduced mass $m$ and momentum $%
\vec{p}=\hbar \vec{k}$ ($\vec{k}$ is the wavevector). Following Ref.~\cite%
{Hamamoto2005}, the adopted interaction potential $V$ consists of three
parts
\begin{eqnarray}
&&V_{\text{cent}}(r)=V_{0}f(r),  \notag \\
&&V_{\text{def}}(\vec{r})=-\beta _{2}k\left( r\right) Y_{20}(\vartheta
,\varphi ),  \label{potential} \\
&&V_{\text{sl}}(r)=-4V_{0}\Lambda ^{2}\frac{1}{r}\frac{df\left( r\right) }{dr%
}(\vec{s}\cdot \vec{l}),  \notag
\end{eqnarray}%
with%
\begin{eqnarray*}
f(r) &=&\frac{1}{1+e^{\frac{r-R}{a}}}, \\
k(r) &=&V_{0}r\frac{df\left( r\right) }{dr}.
\end{eqnarray*}%
Here, $a$, $V_{0}$, and $R$ represent respectively the diffuseness, depth,
and range of the potential with $R=r_{0}A^{1/3}$, where $A$ is mass number
of a nucleus and $r_{0}=1.27$ fm. $\Lambda $ is the reduced Compton
wavelength of nucleon. The solutions of Eq.~(\ref{Hamiltonian}) include the
bound states, resonant states, and the continuum. The bound states can be
obtained by the conventional methods. In order to access the resonant
states, the momentum representation is adopted with the Schr\"{o}dinger
equation as
\begin{equation}
\int d\vec{k^{\prime }}\langle \vec{k}|H|\vec{k^{\prime }}\rangle \psi (\vec{%
k^{\prime }})=E\psi (\vec{k}),  \label{scheq}
\end{equation}%
where $\psi (\vec{k})$ is the momentum wavefunctions. For an axially
deformed system, the parity $\pi $ and the third component of total angular
momentum $m_{j}$ are good quantum numbers. Hence, we can expand the
wavefunctions as
\begin{equation}
\psi (\vec{k})=\psi _{m_{j}}(\vec{k})=\sum_{lj}f^{lj}(k)\phi
_{ljm_{j}}(\Omega _{k}),  \label{wavefunction}
\end{equation}%
where $f^{lj}(k)$ is the radial components of wavefunctions in momentum
representation with the quantum numbers of the orbital angular momentum and
total angular momentum $l$ and $j$. The angular part of wavefunctions is
represented as
\begin{equation}
\phi _{ljm_{j}}(\Omega _{k})=\sum_{m_{s}}\langle lm\frac{1}{2}%
m_{s}|jm_{j}\rangle Y_{lm}(\Omega _{k})\chi _{m_{s}},  \label{spin}
\end{equation}%
where $\chi _{m_{s}}$ is the spin wavefunction with the third component of
spin angular momentum $m_{s}$, and $Y_{lm}(\Omega _{k})$ is the spherical
harmonics in momentum space. Putting the wavefunctions (\ref{wavefunction})
into the equation (\ref{scheq}), the Schr\"{o}dinger equation becomes
\begin{eqnarray}
&&\frac{\hbar ^{2}k^{2}}{2M}f^{lj}(k)+\int k^{\prime }{}^{2}dk^{\prime
}V_{s}(l,j,k,k^{\prime })f^{lj}(k^{\prime })  \notag \\
&&-\beta _{2}\sum_{l^{\prime }j^{\prime }}\int k^{\prime }{}^{2}dk^{\prime
}V_{d}(l,j,l^{\prime },j^{\prime },m_{j},k,k^{\prime })f^{l^{\prime
}j^{\prime }}(k^{\prime })  \notag \\
&=&Ef^{lj}(k),  \label{inteq}
\end{eqnarray}%
with%
\begin{equation*}
V_{s}(l,j,k,k^{\prime })=\frac{2}{\pi }\int r^{2}dr[V_{\text{cent}}(r)+V_{%
\text{sl}}(r)]j_{l}(kr)j_{l}(k^{\prime }r),
\end{equation*}%
and%
\begin{eqnarray*}
&&V_{d}(l,j,l^{\prime },j^{\prime },m_{j},k,k^{\prime }) \\
&=&i^{3l+l^{\prime }}\frac{2}{\pi }\int r^{2}drk\left( r\right)
j_{l}(kr)j_{l^{\prime }}(k^{\prime }r) \\
&&\cdot \sum_{m_{s}}\langle lm|Y_{20}(\Omega _{r})|l^{\prime }m\rangle
\langle lm\frac{1}{2}m_{s}|jm_{j}\rangle \langle l^{\prime }m\frac{1}{2}%
m_{s}|j^{\prime }m_{j}\rangle ,
\end{eqnarray*}%
where $j_{l}(kr)$ and $j_{l^{\prime }}(k^{\prime }r)$ are respectively the
spherical bessel functions of order $l$ and $l^{\prime }$. It is difficult
to solve the integral equation Eq.~(\ref{inteq}). For this reason, we
transform the integral into a sum over a finite set of points $k_{a}$ and $%
dk $ with a set of weights $w_{a}$. And then, the integral equation becomes
a matrix equation
\begin{eqnarray}
&&\frac{\hbar ^{2}k_{a}^{2}}{2M}f^{lj}(k_{a})+%
\sum_{b}w_{b}k_{b}^{2}V_{s}(l,j,k_{a},k_{b})f^{lj}(k_{b})  \notag \\
&&-\beta _{2}\sum_{l^{\prime }j^{\prime
}}\sum_{b}w_{b}k_{b}^{2}V_{d}(l,j,l^{\prime },j^{\prime
},m_{j},k_{a},k_{b})f^{l^{\prime }j^{\prime }}(k_{b})  \notag \\
&=&Ef^{lj}(k_{a}).  \label{sumeq}
\end{eqnarray}%
The Hamiltonian matrix in Eq.~(\ref{sumeq}) is not symmetrical. For
simplicity in computation, we symmetrize the Hamiltonian matrix by the
transformation
\begin{equation}
f(k_{a})=\mathbf{f}(k_{a})/\sqrt{w_{a}}k_{a}.  \label{sys}
\end{equation}%
Putting Eq.~(\ref{sys}) into Eq.~(\ref{sumeq}), we obtain a symmetric matrix
equation in the momentum representation as
\begin{eqnarray}
&&\frac{\hbar ^{2}k_{a}^{2}}{2M}\mathbf{f}^{lj}(k_{a})+\sum_{b}\sqrt{%
w_{a}w_{b}}k_{a}k_{b}V_{s}(l,j,k_{a},k_{b})\mathbf{f}^{lj}(k_{b})  \notag \\
&&-\beta _{2}\sum_{b}\sum_{l^{\prime }j^{\prime }}\sqrt{w_{a}w_{b}}%
k_{a}k_{b}V_{d}(l,j,l^{\prime },j^{\prime },m_{j},k_{a},k_{b})\mathbf{f}%
^{l^{\prime }j^{\prime }}(k_{b})  \notag \\
&=&E\mathbf{f}^{lj}(k_{a}).  \label{simeq}
\end{eqnarray}%
So far, to solve the Schr\"{o}dinger equation becomes an eigensolution
problem of the symmetric matrix. All the bound states and resonant states
can be acquired simultaneously by diagonalizing the Hamiltonian in Eq.~(\ref%
{simeq}). The diagonalization of the Hamiltonian matrix in Eq.~(\ref{simeq})
can give us the energies and wavefunctions in the momentum space. To obtain
the wavefunctions in the coordinate space, the following transformation is
adopted
\begin{equation}
\psi \left( \vec{r}\right) =\psi _{m_{j}}(\vec{r})=\frac{1}{\left( 2\pi
\right) ^{3/2}}\int d\vec{k}e^{i\vec{k}\cdot \vec{r}}\psi _{m_{j}}(\vec{k}).
\label{Rwavefunctions}
\end{equation}%
Assuming that the $\psi _{m_{j}}(\vec{r})$ holds the form
\begin{equation}
\psi _{m_{j}}(\vec{r})=\sum_{lj}f^{lj}(r)\phi _{ljm_{j}}(\Omega _{r}),
\end{equation}%
we can obtain the radial part of wavefunctions
\begin{equation}
f^{lj}(r)=i^{l}\sqrt{\frac{2}{\pi }}\sum_{a=1}^{N}\sqrt{w_{a}}%
k_{a}j_{l}(k_{a}r)\mathbf{f}^{lj}(k_{a}),
\end{equation}%
and the radial density distributions%
\begin{equation}
\rho _{m_{j}}(r)=\sum_{lj}f^{lj\ast }(r)f^{lj}(r).
\end{equation}

\section{numerical details and results}

With the formalism represented above, we explore the single-particle
resonances in $^{31}$Ne in order to clarify the physical mechanism of $^{31}$%
Ne halo. Before starting the researches, we introduce numerical details. In
the present calculations, the coupled-channel method is adopted to solve the
equation of motion describing deformed nuclei, and the wavefunctions are
expanded with different channels labeled as $lj$. The sum over $lj$ in Eq.~(%
\ref{wavefunction}) is restricted to a limited range $N_c=8$. In Eq.~(\ref%
{inteq}), the momentum integration from zero to infinite is truncated to a
large enough $k_{max}=4.0$ fm$^{-1}$, and replaced by a sum using the
Gauss-Legendre quadrature with a finite grid number $N_l=120$. The choice of
these parameter values is enough to ensure the required precision.
\begin{figure}[tph]
\centering%[!ht]
\includegraphics[width=8.cm]{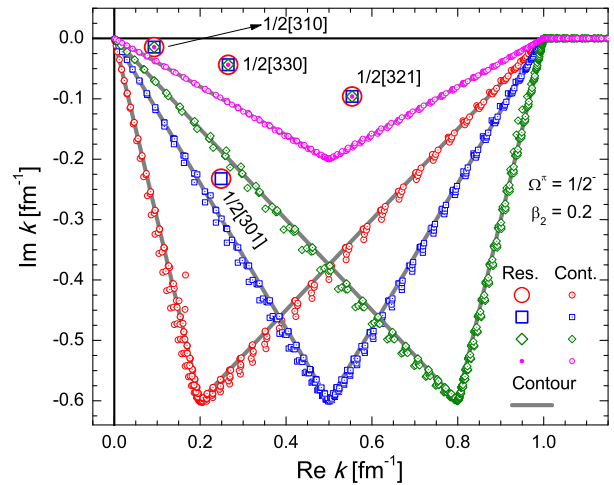}\centering
\caption{(Color online) Single-particle spectra for the states $\Omega^{%
\protect\pi}=1/2^{-}$ with $\protect\beta_{2}=0.2$ in complex momentum
plane, where the red open circles, blue hollow squares, green hollow
rhombus, and magenta points represent the resonant states in four different
contours, respectively. The other is the continuum following the contours in
grey solid line.}
\end{figure}
\begin{figure}[tph]
\centering%[!ht]
\includegraphics[width=8.cm]{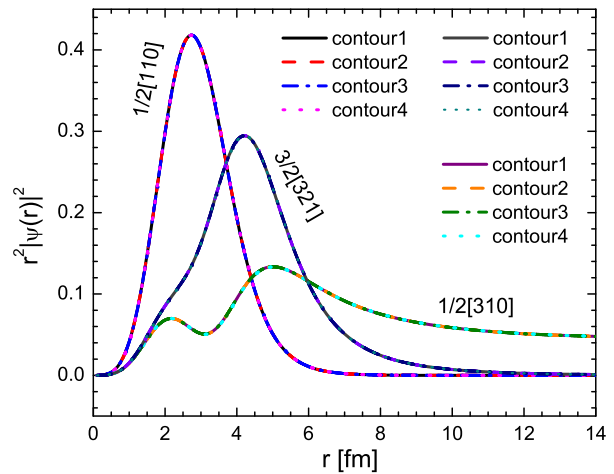}\centering
\caption{(Color online) Radial density distributions as a function of the
coordinate $r$ for the single-particle states 1/2[110] and 1/2[310] with $%
\protect\beta_{2}$ = 0.2 and 3/2[321] with $\protect\beta_{2}$ = 0.5. The
left four lines belong to 1/2[110], the upper four lines on the right belong
to 3/2[321], and those in the lower belong to 1/2[310]. The four contours
are the same as those in Fig.~1.}
\end{figure}

With these parameters designed, we check the validity and applicability of
present calculations. The numerical results for the states $\Omega ^{\pi
}=1/2^{-}$ with the deformation $\beta _{2}=0.2$ are shown in Fig.~1, where
the four different contours are adopted in the momentum integration. The
resonant states are completely independent on the contours. With the contour
becomes deeper from the magenta color to the blue color and/or the contour
from left (red color) to right (green color), the resonant states always
stay at their original locations in the complex momentum plane. The resonant
state 1/2[301] is located outside the magenta and green contours. Hence, it
is necessary to select a large enough contour to expose all the resonant
states concerned.

In addition to the single particle energies, the wavefunctions are also
independent of contours. In Fig.~2, we have plotted the radial density
distributions as a function of the coordinate $r$ for the single-particle
states 1/2[110], 1/2[310], and 3/2[321]. The state 1/2[110] is deeply
bound over the range of deformation under consideration. The state 1/2[310]
locates at a resonant orbit with $\beta_2=0.2$. The state 3/2[321] is weakly bound
in the position of $\beta_2=0.5$. Whether the bound state, the resonant state, or the weakly bound state, the calculated density distributions with the four different contours
completely coincide together, which indicates the wavefunctions are
independent of contours. Compared with the bound state 1/2[110], the radial
density distributions for the resonant state 1/2[310] and the weakly bound
state 3/2[321] spatially extend in a larger range, which indicates that a
halo may be formed if a nucleon populates in the orbits.

\begin{figure}[tph]
\centering
\includegraphics[width=8.0cm]{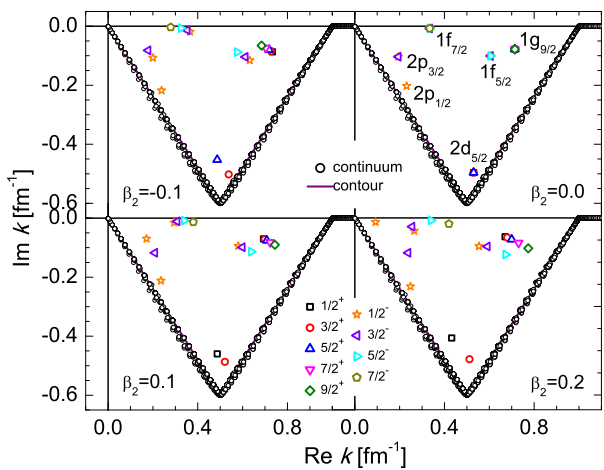}
\caption{ (Color online) Single-particle spectra of $^{31}$Ne for the states
$\Omega^{\protect\pi}=1/2^{\pm}, 3/2^{\pm}, \cdots, 9/2^{\pm}$ in several
different deformations. The black open circles and purple solid line
represent respectively the continuum and contour of momentum integration,
and the other represent the resonant states.}
\end{figure}
\begin{figure}[tph]
\centering%[!ht]
\includegraphics[width=8.cm]{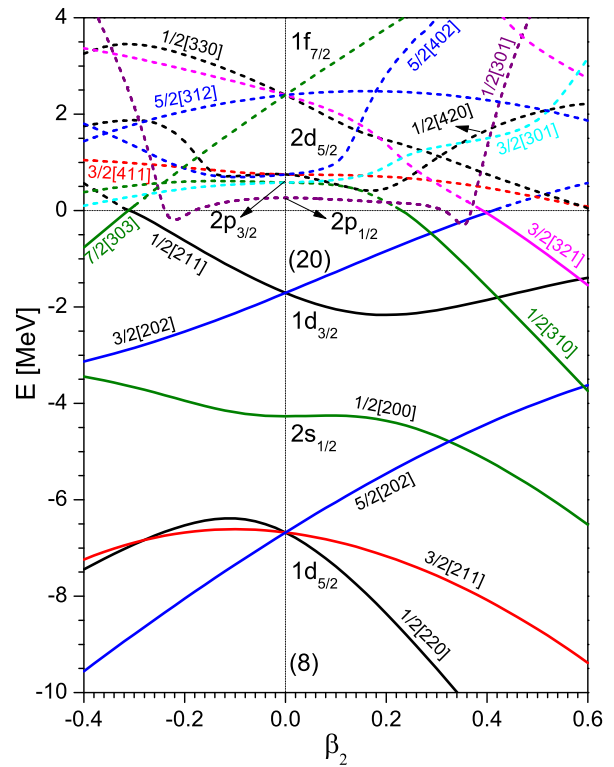}\centering
\caption{(Color online) Neutron single-particle levels as a function of
quadruple deformation $\protect\beta_{2}$. The bound levels and resonant
levels are marked by the solid line and dashed line, respectively.}
\end{figure}

As the resonant states are independent of the contours, we can select a
large enough contour to expose all the resonant states concerned. Using the
triangle contour with the four points $k=0$ fm$^{-1}$, $k=0.5-i0.6$ fm$^{-1}$%
, $k=1.0$ fm$^{-1}$, and $k=4.0$ fm$^{-1}$, these resonant states concerned
are obtained over the range of deformation under consideration. In Fig.~3,
we have presented the solutions of Eq.~(\ref{simeq}) for the states $\Omega
^{\pi }=1/2^{\pm },3/2^{\pm },\cdots ,9/2^{\pm }$ in several different
deformations. It is noticed that the positions of the resonant states in
complex momentum plane depend on the deformation $\beta _{2}$. When the
spherical symmetry is broken ($\beta _{2}\neq 0$), the states $2p_{3/2}$, $%
1f_{5/2}$, $1f_{7/2}$ and $1g_{7/2}$ are split into two, three, four, and
five resonant states, respectively. Similarly, the state $2d_{5/2}$ is split
into three resonant states. When $\beta _{2}=-0.1$, only the $5/2^{+}$ and $%
3/2^{+}$ states emerge in the complex momentum plane, while the $1/2^{+}$
state is beyond the range of contour. When $\beta _{2}=0.1$ and $\beta
_{2}=0.2$, only the $1/2^{+}$ and $3/2^{+}$ states appear in the complex
momentum plane, while the $5/2^{+}$ state lies outside of the contour. These
show that some resonances may move into the contour and some other
resonances may be removed from the contour with the change of deformation.

Making clear the dependence of the resonance on deformation, let's explore
the mechanism of deformation halo in $^{31}$Ne. The calculated
single-particle energies for all the bound and resonant states concerned are
shown in Fig.~4, where the bound (resonant) levels are marked by the solid
(dashed) line with the Nilsson labels on the line. The corresponding
spherical labels are marked in the position $\beta _{2}=0.0$. From Fig.~4,
it can be seen that the bound levels are the fully same as those in Refs.~%
\cite{Hamamoto2010,Liu2012}. For the resonant levels, only those with the
spherical labels $1f_{7/2}$ are obtained in the scattering phase shift
calculations~\cite{Hamamoto2010}. Although the more resonant levels
including those with the spherical labels $1f_{7/2}$, $1f_{5/2}$, and $%
1g_{9/2}$ are obtained in the complex scaled calculations~\cite{Liu2012},
these resonant levels with the spherical labels $2p_{3/2}$, $2p_{1/2}$, and $%
2d_{5/2}$ are still missed. Since the halo is mainly attributed to an
occupation of the orbit with lower angular momentum and separation energy by
valance nucleon, these resonant levels with the spherical labels $2p_{3/2}$,
$2p_{1/2}$, and $2d_{5/2}$ are more attractive. Unfortunately, these
resonant levels have not been obtained in the scattering phase shift
calculations and the complex scaling calculations. Furthermore, the levels
1/2[310] and 1/2[330] obtained in the scattering phase shift is
questionable. With the decrease of deformation from $\beta _{2}=0.6$ to $%
\beta _{2}=0.0$, the 1/2[310] level evolves from a bound state to a resonant
state. At $\beta _{2}=0.0$, the 1/2[310] level degenerates the $2p_{3/2}$
level rather than the $1f_{7/2}$ level. Over the range of deformation, the
1/2[330] level is always located in the resonant region, while it is
incorrectly connected to the level 1/2[310] in Ref.~\cite{Hamamoto2010}.
With these resonant levels in the spherical labels $2p_{3/2}$, $2p_{1/2}$,
and $2d_{5/2}$ obtained, an inverted ordering of $p-f$ orbits is discovered
in comparison with the usual shell model prediction. As it happens the $p-f$
inversion, the $21$st neutron in $^{31}$Ne can occupy a $p-$orbit or that
with major $p-$component, which is the cause of the halo formed in $^{31}$%
Ne. Assuming that $^{31}$Ne is a prolate nucleus, the $21$st neutron may
occupy the levels 1/2[301], 1/2[310], 3/2[202], 3/2[321], and 1/2[211] in
the intervals of $0.0\leqslant \beta _{2}\leqslant 0.20$, $0.20\leqslant
\beta _{2}\leqslant 0.29$, $0.29\leqslant \beta _{2}\leqslant 0.40$, $%
0.40\leqslant \beta _{2}\leqslant 0.59$, and $\beta _{2}\geqslant 0.59$,
respectively. Which orbit is the most appropriate for $^{31}$Ne? The state
1/2[301] locates in the resonant level. If a nucleon occupies the level, its
density distribution can spatially extend far beyond the nuclear core.
However, there exists a large energy gap in the interval of $0.0\leqslant
\beta _{2}\leqslant 0.20$, which increases the difficulty of nucleon
occupying the orbit. According to the claim in Ref.~\cite{Sagawa1992}, the
halo structure is ascribed to an occupation of the $l=0$ or $l=1$ orbit by
valence nucleon with lower separation energy. From this view, the
appropriate orbit may be 1/2[310] in the interval of $0.20\leqslant \beta
_{2}\leqslant 0.29$ or 3/2[321] in the interval $0.40\leqslant \beta
_{2}\leqslant 0.59$. The result agrees with the calculations in Ref.~\cite{Urata2011,Urata2012}, where the ground state of $^{31}$Ne is predicted with the configurations of $I^{\pi}= 3/2^{-}$ at $\beta
_{2}=0.2$ or $0.55$. In Ref.~\cite%
{Minomo2011,Minomo2012}, Minomo $et$ $al$.
concluded that neutron-rich Ne isotopes are strongly deformed and $^{31}$Ne
has a halo structure with the spin parity $3/2^{-}$. The similar conclusion is obtained in Ref.~\cite{Sumi2012} for $^{31}$Ne with the deformation $\beta _{2}\sim 0.4$. In addition, an appropriate
configuration 3/2[321] is predicted with the deformation in the interval $0.40<\beta _{2}<0.59
$ for the ground state of $^{31}$Ne in Ref.~\cite{Hamamoto2010}. These conclusions are in agreement with our calculations with the inclination of the orbit 3/2[321] in large deformation.

\begin{figure}[tph]
\centering%[!ht]
\includegraphics[width=8.cm]{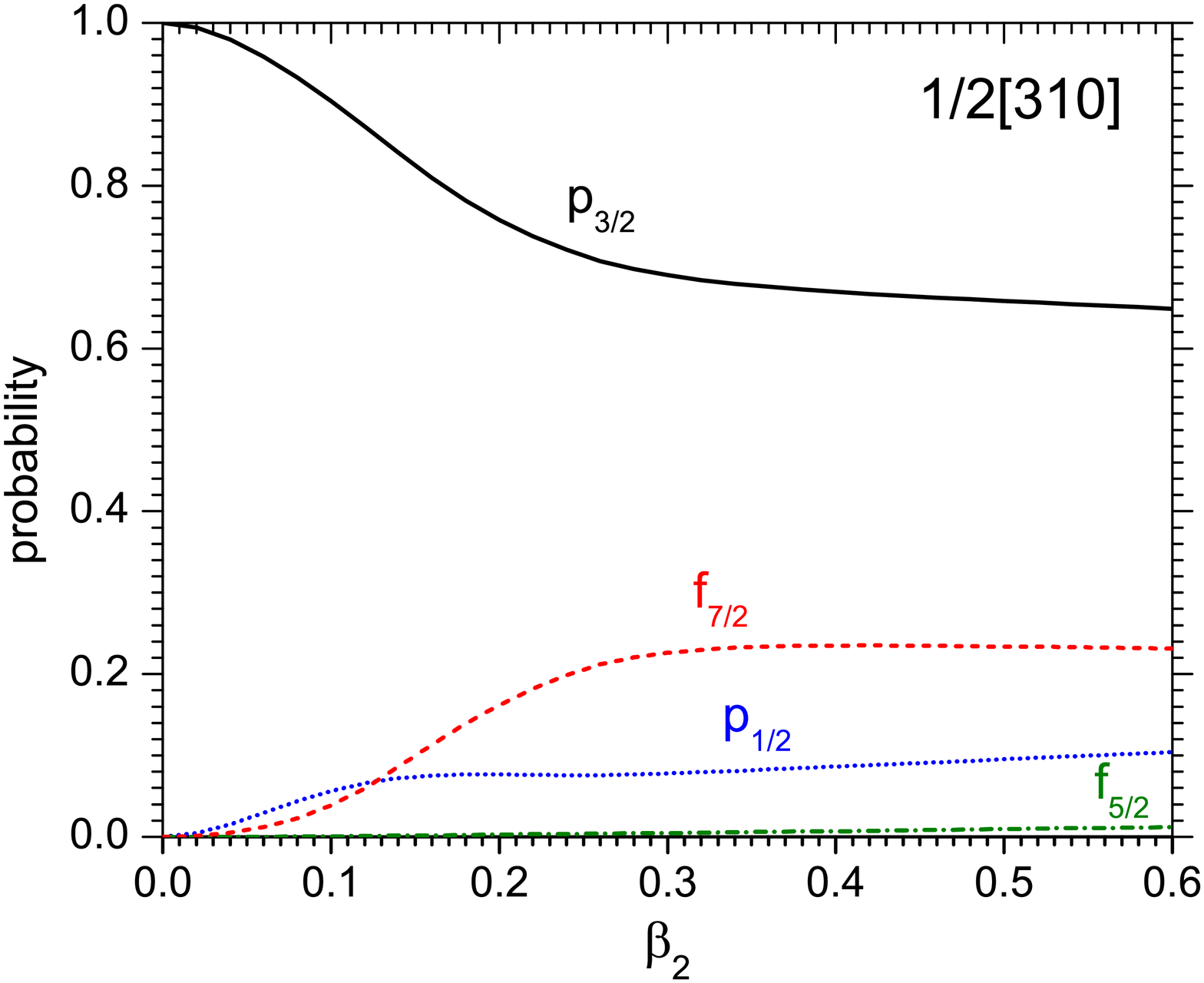}\centering
\caption{(Color online) Calculated occupation probabilities of major
components in the 1/2[310] level, where the black solid line, red dashed
line, blue dotted line, and green dash-dotted line represent the $p_{3/2}$, $%
f_{7/2}$, $p_{1/2}$, and $f_{5/2}$ components, respectively.}
\end{figure}

In order to identify the most likely orbit in favor of halo in $^{31}$Ne, we
compare the occupation probabilities of major components in the levels
1/2[310] and 3/2[321]. The calculated occupation probabilities of major
components as a function of deformation $\beta _{2}$ are plotted in Fig.~5
for the level 1/2[310]. Over the range of deformation, the occupation
probabilities of the four components $p_{3/2}$, $f_{7/2}$, $p_{1/2}$, and $%
f_{5/2}$ are observable, while the other components are negligible. At $%
\beta _{2}=0.0$, the level 1/2[310] degenerates the spherical configuration $%
2p_{3/2}$, which supports the result from the level 1/2[310] in Fig.~4. With
the increase of deformation, the occupation probability of $p_{3/2}$
decreases, while the occupation probabilities of $f_{7/2}$, $p_{1/2}$, and $%
f_{5/2}$ increase. Even so, the wavefunction of the level 1/2[310] is mostly
constituted by the $p_{3/2}$ configuration. Within the range of deformation,
the occupation probability of $p_{3/2}$ is over 64\%, which indicates that
the level 1/2[310] satisfies the condition of halo formation and the halo in
$^{31}$Ne may come from an occupation of the orbit 1/2[310] by valance
neutron.
\begin{figure}[tph]
\centering%[!ht]
\includegraphics[width=8.cm]{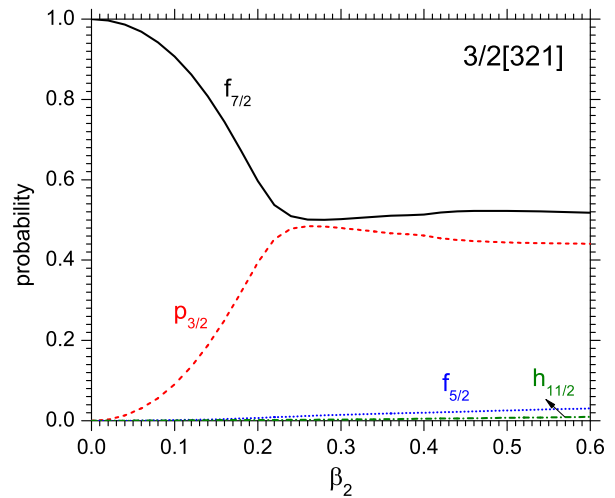}\centering
\caption{(Color online) Calculated occupation probabilities of major
components in the 3/2[321] level, where the black solid line, red dashed
line, blue dotted line, and green dash-dotted line represent the $f_{7/2}$, $%
p_{3/2}$, $f_{5/2}$, and $h_{11/2}$ components, respectively.}
\end{figure}

For comparison, we have also plotted the occupation probabilities of major
components in the wavefunction of the level 3/2[321] as a function of $\beta
_{2}$ in Fig.~6. Similar to the level 1/2[310], only the four components $%
f_{7/2}$, $p_{3/2}$, $f_{5/2}$, and $h_{11/2}$ are observable, while the
other components are insignificant. At $\beta _{2}=0$, the level 3/2[321]
degenerates the spherical configuration $1f_{7/2}$, which agrees the case in
Fig.~4. With the increase of deformation from $\beta _{2}=0$ to $\beta
_{2}=0.26$, the occupation probability of $f_{7/2}$ decreases, while the
occupation probabilities of $p_{3/2}$, $f_{5/2}$, and $h_{11/2}$ increase.
Further increasing the deformation, the occupation probabilities of $f_{7/2}$
and $p_{3/2}$ do almost unchange and the wavefunction of the level 3/2[321]
is mostly constituted by the $f_{7/2}$ and $p_{3/2}$ configurations.
As the
occupation probability of $p_{3/2}$ in the wavefunction of the level
3/2[321] is over 44\%, the level 3/2[321] also supports the formation of
halo. However, the energy gap between the levels 3/2[321] and 1/2[211] is
considerably large in the interval $0.40\leqslant \beta _{2}\leqslant 0.50$.
If nuclear deformation is larger than $\beta _{2}=0.50$, the energy gap
between the levels 3/2[321] and 1/2[211] becomes small, the halo in $^{31}$%
Ne is most likely derived from an occupation of the orbit 3/2[321] by a
valance neutron. Nevertheless, the large contamination of
$f$-wave in 3/2[321] implies that the core excited component $
^{30}$Ne(2$^+$)$\otimes$ f$_{7/2}$ is non-negligible, if the ground state is
3/2$^-$. 
\begin{figure}[tph]
\centering%[!ht]
\includegraphics[width=8.cm]{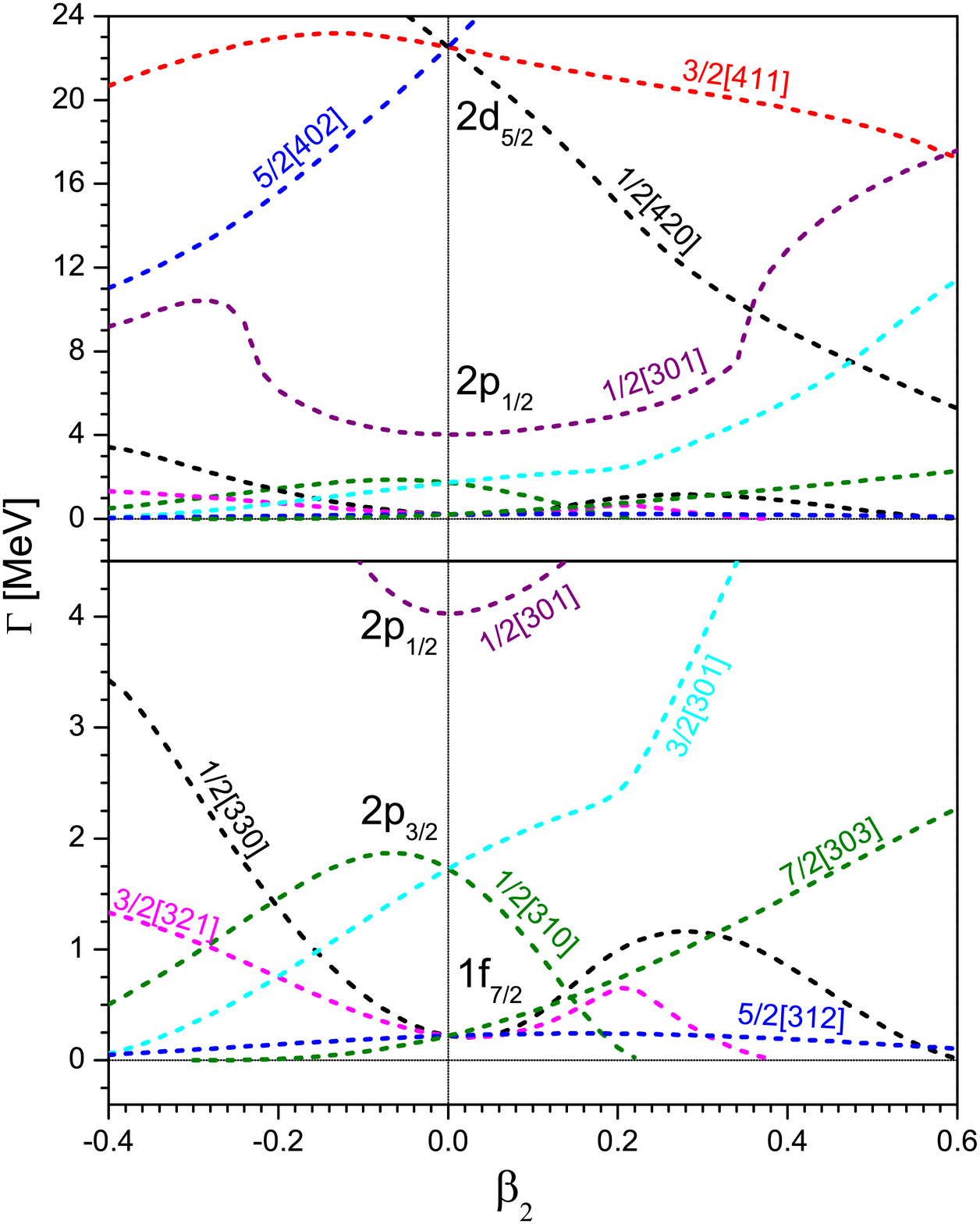}\centering
\caption{(Color online) The same as Fig.~4, but for the width of resonant
states.}
\end{figure}

With the knowledge on the energy levels (positions) of resonant states,
let's figure out the widths of resonant states. The widths as a function of
deformation $\beta _{2}$ is drawn in the upper panel in Fig.~7 for all the
resonant states concerned. The lower panel in Fig.~7 is a close-up of the
widths for these resonant states near zero energy surface. For simplicity in
comparison, the marks of the widths in Fig.~7 are the fully same as those of
the levels shown in Fig.~4. Similar to the levels, there exists the shell
structure in the widths. It is different where appears the energy gap. For
the levels, a large energy gap appears between the states $1f_{7/2}$ and $%
2d_{5/2}$ at $\beta _{2}=0.0$. However for the widths, an unexpected gap
appears between the states $2d_{5/2}$ and $2p_{1/2}$ at $\beta _{2}=0.0$. In
addition, the order of the widths is different from that of the levels. The
resonant level $2d_{5/2}$ lies between the levels $1f_{7/2}$ and $2p_{3/2}$,
while the corresponding width is located above them. Due to the lower
centrifugal barrier, the width of a $p-$state is usually greater than that
of a $f-$state, which can be seen in the lower panel of Fig.~7. With the
increase of deformation, the level $2p_{3/2}$ is split into two levels
1/2[310] and 3/2[301]. The resonant state 1/2[310] becomes more stable with
the increase of deformation from $\beta _{2}=0.0$, while the states 3/2[301]
and 1/2[301] become more unstable. Hence, it is relatively easy to form halo
in the resonant state 1/2[310] in $^{31}$Ne. Similar phenomena also emerge
in the $f-$state. With the increase of deformation, the level $1f_{7/2}$ is
split into four levels 1/2[330], 3/2[321], 5/2[312], and 7/2[303]. The
resonant state 3/2[321] becomes more unstable with the increase of
deformation from $\beta _{2}=0.0$ to $\beta _{2}=0.2$. Further increasing
deformation, the resonant state 3/2[321] becomes more stable, which implies
that the state 3/2[321] also supports the formation of halo in $^{31}$Ne in
the region of large deformation. The conclusion is in agreement with that
obtained in terms of the single-particle levels. All these show that $^{31}$%
Ne is a halo nucleus formed by $p_{3/2}$ configuration.

In order to provide the possible observables for experiment, in Fig.~8 we have plotted the radial-momentum probability distributions for the state 1/2[310] with $\beta_2=0.25$ and 3/2[321] with $\beta_2=0.5$. The remarkable differences in the position and height of peak and the range of radial-momentum probability distributions can be identified by the knock-out reaction, which is helpful to distinguish the two orbits.

\begin{figure}[tph]
\centering%[!ht]
\includegraphics[width=8.cm]{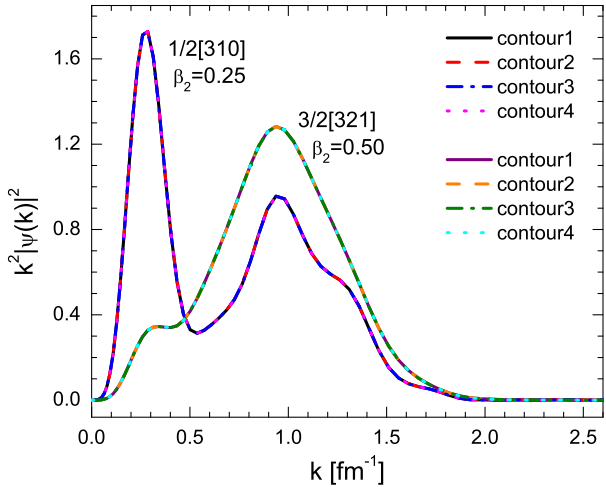}\centering
\caption{(Color online) Radial-momentum probability distributions for the state 1/2[310] with $\beta_2=0.25$ and 3/2[321] with $\beta_2=0.5$, where the upper and lower four lines belong to 1/2[310] and 3/2[321], respectively.}
\end{figure}

\section{Summary}

In summary, the complex momentum representation method is used to explore
the physical mechanism of deformed halo in $^{31}$Ne with the theoretical
formalism presented. The Schr\"{o}dinger equation describing deformed nuclei
is processed into a set of coupled differential equations by the
coupled-channel method. This set of coupled differential equations is solved
using the complex momentum representation technique, which makes the
solutions of the Schr\"{o}dinger equation become the diagonalization of a
matrix, and the bound and resonant states are treated on the equaling
footing. The dependence of the calculations on the unphysical parameters is
checked, satisfactory result is obtained.

The single-particle resonant states in $^{31}$Ne are explored. Not only the
narrow resonance $1f_{7/2}$ but also the broad resonances $2p_{3/2}$, $%
2p_{1/2}$, and $2d_{5/2}$, which are missed in the scattering phase shift
and the complex scaling calculations, are obtained. As these broad
resonances come into sight, the $p-f$ inversion is seen clearly in the
single-particle levels. As a result, the $21$st neutron in $^{31}$Ne can
occupy a $p-$orbit or that with major $p-$component, which is the cause of
the halo formed in $^{31}$Ne. From the position and energy gap, it can be
judged that two appropriate Nilsson orbits are 1/2[310] in the interval of $%
0.20\leqslant \beta _{2}\leqslant 0.29$ and 3/2[321] in the interval $%
0.40\leqslant \beta _{2}\leqslant 0.59$. Over the range of deformation, the
wavefunction of the level 1/2[310] is mainly constituted by the $p_{3/2}$
configuration, which supports the formation of halo. When $\beta
_{2}\geqslant 0.40$, the occupation probability of $p_{3/2}$ in the state
3/2[321] is over 44\%, which also supports the formation of halo. All these
show that $^{31}$Ne is a halo nucleus formed by the $p_{3/2}$ configuration.
This conclusion is also supported by the calculated widths.

\begin{acknowledgments}
This work was partly supported by the national natural science foundation of
China (11575002, 11175001, and 11305002); the program for new century
excellent talents in university of China (NCET-05-0558); the foundation of
academic and technical leaders in Anhui province (J05010247), and the Doctoral Scientific Research Startup Fund of Anhui University (J01001319-J10113190082).
\end{acknowledgments}

\end{document}